\documentstyle[amstex,aps,eqsecnum]{revtex}
\def\A{{\cal A}}                            
\def\B{{\cal B}} 
 
\def\G{{\cal G}}                            
\def\L{{\cal L}}                            
\def\M{{\cal M}}                            
\def\bds#1{\boldsymbol{#1}}

\begin{document} 
\title{Spectral partitions on infinite graphs}
\author{Raffaella Burioni, Davide Cassi}
\address{ Dipartimento di Fisica and Istituto Nazionale di Fisica della Materia, \\
           Parco Area delle Scienze 7/A, 43100 Parma, Italy 
              }
\author{Claudio Destri}
\address{  Dipartimento di Fisica G. Occhialini,  and INFN, sezione di Milano,\\
          Via Celoria 16,  20133 Milano, Italy \vskip 1 truecm 
              }
\date{\today}
\maketitle
\begin{abstract} 
Statistical models on infinite graphs may exhibit inhomogeneous
thermodynamic behaviour at macroscopic scales. This phenomenon is of 
geometrical origin and may be properly described in terms of {\em spectral
partitions} into subgraphs with well defined spectral dimensions and
spectral weights. These subgraphs are shown to be thermodynamically
homogeneous and effectively decoupled.
\end{abstract}



\section{Introduction}
The study of model systems without translation invariance is
an interesting and complex subject of modern statistical mechanics.
A very general description of this situation is in terms of
statistical models on graphs, that is on generic networks
formed by sites, where dynamical variables reside, and links 
connecting pairwise sites whose variables are coupled. This is the 
direct extension of the typical setup valid for crystalline lattices,
which are indeed very special, homogeneous graphs.

On the other hand, graphs are not in general homogeneous and the main
question is how these inhomogeneities affect physical properties and
give rise to relevant changes with respect to lattices. While small
scale inhomogeneities will affect local properties, one expects that
only large scale inhomogeneities are relevant for bulk thermodynamic
properties. Most likely, the latter properties are those that show
universal features which depend only on a few global parameters, just
as in the case of lattices. The study of such universality requires
consideration of infinite graphs (with certain natural restrictions given
below), where the thermodynamic limit is taken. 
 
The main relevant geometrical parameter affecting universal properties
is the spectral dimension $\bar d$ of an infinite graph $\G$
\cite{ao,hhw,debole}.  It generalizes the Euclidean dimension of
lattices to arbitrary real values and is naturally defined from the
infrared behaviour of the spectral density of the Laplacian operator on
$\G$ \cite{debole}.  An equivalent definition, the one adopted in this work,
is in terms of average properties of random walks on $\G$ at large times, that
is to say of the singularities of the Gaussian model on the same graph
\cite{hhw,debole}.

On the other hand, the spectral dimension of the whole graph $\G$, by
itself turns out not to be sensitive to macroscopic inhomogeneities
strong enough to give rise to true thermodynamic inhomogeneities.
Indeed it may happen that distinct macroscopic parts of an infinite
graph exhibit distinct thermodynamic behaviour. We shall show below
that such parts can be characterized in terms of their own spectral
dimension, possibly plus a spectral weight, resulting in an effective
{\em spectral partition} of $\G$. The crucial point is that these
parts form subgraphs which are thermodynamically {\em independent},
that is to say completely uncoupled as far as thermodynamic properties
are concerned. In other words, inhomogeneous thermodynamic behaviour
on the same infinite graph necessarily imply effective decoupling.

\section{Infinite graphs: basic definitions, measure and averages}

A (unoriented) graph $\G$ \cite{harari} is the ordered couple
$(G\,,G_L)$ formed by a countable set $G$ of vertices (or sites, or
nodes), that we shall generically indicate with small-case Latin
letters, $i$, $j$, $k$, \ldots, and a set $G_L$ of unoriented links
(or bonds) which connect pairwise the sites and are therefore
naturally denoted by couples $(i,j)=(j,i)$. When the set $G$ is
finite, $\G$ is a {\em finite} graph and we shall denote $N$ the
number of vertices of $\G$. A subgraph $\G'$ of $\G$ is a graph such
that $G' \subseteq G$ and $G_L' \subseteq G_L$. A subgraph is said to
be {\em complete} if its has all the available links, that is if, given
the subset of nodes $G'$, the subset of links $G_L'$ is the largest 
possible one.

A path in $\G$ is a sequence of consecutive links
$\{(i,k)(k,h)\dots(n,m)(m,j)\}$.  A graph is said to be connected, if
for any two points $i,j \in G$ there is always a path joining
them. In the following we will consider only connected graphs.

The graph topology can be algebraically described by  
its adjacency matrix $\bds A$ with elements
\begin{equation}
A_{ij}=\left\{
\begin{array}{cl}
1 & {\rm if } \ (i,j) \in G_L \\
0 & {\rm if } \ (i,j) \not\in G_L \\
\end{array}
\right .
\label{defA}
\end{equation}
The Laplacian  matrix $\bds L$ on the graph $\G$ has elements:
\begin{equation}
L_{ij} = z_i \delta_{ij} - A_{ij}
\label{defDelta}
\end{equation}
where $z_i=\sum_j A_{ij}$, the number of nearest neighbours of $i$, 
is called the coordination number (or degree) of site $i$. Here we will 
consider graphs with $z_{max}=\sup_{i} z_i < \infty$. 

One can also consider a generalization of the adjacency matrix, which
corresponds to the ferromagnetic and uniformly bounded
coupling $J_{ij}$, with $J_{ij} \neq 0 \iff A_{ij}=1$ and
$\sup J_{ij} < \infty$, $\inf J_{ij} > 0$. 
The elements of the generalized Laplacian matrix then read:
\begin{equation}
\L_{ij} = J_i \delta_{ij} - J_{ij}
\label{defL}
\end{equation}
where $ J_i =\sum_j J_{ij}$.

Every connected graph $\G$ is endowed with an intrinsic metric 
generated by the chemical distance $r_{i,j}$  
which is defined as the number of links in the shortest path(s) 
connecting vertices $i$ and $j$. 

Let us now consider thermodynamic averages on infinite graphs
\cite{woess}. The Van Hove sphere $S_{o,r} \subset \G$ of centre $o\in
G$ and radius $r$ is the complete subgraph of $\G$ containing all
$i\in G$ whose distance from $o$ is $\le r$ and all the links of $\G$
joining them. We will call $N_{o,r}$ the number of vertices contained
in $S_{o,r}$.

In the thermodynamic limit the average $[f]_G$ of a 
real--valued function $f$ on $G$ is:
\begin{equation}
[f]_G\equiv 
\lim_{r\rightarrow\infty} 
{\displaystyle 1 \over \displaystyle N_{o,r}}\sum_{i\in S_{o,r}} f_i
\label{deftd}
\end{equation}
This average does not depend on the choice of the origin $o\in G$
provided  $f$ is bounded from below and  
\begin{equation}
\lim_{r\to \infty} {|\partial S_{o,r}|\over N_{o,r}}~=~0
\label{isop} 
\end{equation}
where $|\partial S_{o,r}|$ is the number of the vertices of the sphere 
$S_{o,r}$ connected with the rest of the graph \cite{rimtim}. 
Here we shall restrict our attention to graphs with this property.

The measure $|A|$ of a subset $A \subset G$ is the average value 
$[\chi(A)]_G$ of its characteristic function $\chi_i(A)$ defined by
$\chi_i(A)=1$ if $i\in A$ and $\chi_i(A)=0$ if $i\not\in A$.
The measure of a subset of links $G_L'\subseteq G_L$ is similarly given by:
\begin{equation}
|G_L'| \equiv 
\lim_{r\rightarrow\infty} 
{\displaystyle N'_{L,r} \over  N_{o,r}}
\label{mislinks}
\end{equation}
where $N'_{L,r}$ is the number of links of $G_L'$ contained in the
sphere $S_{o,r}$.  Any two nonzero--measure subsets $A$ and $B$ of $G$
are said to be equivalent if their symmetric difference has zero
measure, that is $|A|=|B|=|A\cap B|$. For any given nonzero--measure
subsets $A\subset G$ we shall denote $\{A\}$ its equivalence class. 
Then $A$ is said to be a representative of $\{A\}$. With the
subgraph $\G'$ defined by the ordered double $(G'\,,G_L')$, we
identify the measure of the subgraph as the measure $|G'|$ of its
points.

Given a (nonzero--measure) subset $A\subset G$, we define the average on $A$ of
any real--valued function $f$ on $G$ as:
\begin{equation}
	[f]_A = [\chi(A)\,f]_G
\label{mediaA}
\end{equation} 
By definition $[f]_A$ is a function only of the equivalence classes,
that is $[f]_A=[f]_{\{A\}}$. Moreover, quite evidently
$[f]_C=[f]_A+[f]_B$ whenever $C=A\cup B$ and $|A\cap B|=0$.

Given a complete subgraph $\M=(M\,,M_L)$, we denote ${\bar \M}$
its complement in $\G$. 
This is formed by all points that do not belong to $M$ and
by all links of $\G_L$ which connect them. ${\bar \M}$ is therefore a
complete subgraph. We call the pair $(\M\,,{\bar \M})$ a partition of
order two of $\G$ whenever both $M$ and its complement ${\bar M}$ are
nonzero--measure subsets of $G$. 

We introduce now the important concept of 
{\em minimal distance} $\underline{D}(\A,\B)$ between 
any pair $\A,\B$ of nonzero--measure
subgraphs of $\G$ such that $|A\cap B|=0$. It is defined as
\begin{equation}
	\underline{D}(\A,\B) = \min \bigl( n \;:\; 
	\bigl| A \,{\cap}_n B\bigr| >0  \bigl)
\end{equation}
where
\begin{equation}
	A \,{\cap}_n B = \{i\in A\;:\; \text{dist}(i,B) = n\}
	\;,\quad \text{dist}(i,B) = \min_{j\in B} \text{r}_{(i,j)}
\end{equation}
For $n=0$, ${\cap}_n$ reduces to the usual intersection operator.
Notice that, while in general the relation $A\,{\cap}_n B$ is not
symmetric in $A,B$, the minimal distance is symmetric:
$\underline{D}(\A,\B)=\underline{D}(\B,\A)$.  In fact, from the
boundedness of $z_i$, it can be shown by induction on $n$ that
\begin{equation}
	|B\cap_n A| \ge  (z_{max})^{-n}\, |A \cap_n B|
\end{equation}
so that  
\begin{equation}
	|A\cap_n B| > 0 \Longrightarrow |B \cap_n A|>0
\end{equation}
implying our assertion.

Consider now the minimal distance between the two members of a
partition of order two. Suppose $\underline{D}(\M,\bar \M)=n>1$; then
$|M \cap_n \bar M|>0 \Rightarrow |M\cap_{n-1} \bar M|>0$ from the
boundedness of $z_i$. This implies that if $\underline{D}(\M,\bar \M)$
is finite, then $\underline{D}(\M,\bar \M)=1$.  In this case we may
say that $\M$ and $\bar \M$ are {\em densely interlaced}, while in the
opposite case that they are {\em infinitely separated}. From the
definition of minimal distance, it follows that if two subgraphs $\A$
and $\B$ of $\G$ are infinitely separated, their common frontier
$\partial (\A,\B)$ (i.e. the links $(i,j)\in \G_L$ with $i \in A$ and
$j\in B$) is a zero--measure set. Then the two subgraphs can be
disconnected by cutting such a zero--measure set of links. This relates
the property of infinite separability to the simple separability
property defined in \cite{rimtim}. Indeed, the two definitions
coincide. We shall term {\em separable partition} a partition
$(\M\,,{\bar \M})$ where $\M$ and $\bar \M$ are infinitely separated.

\section{The Gaussian model: infrared behaviour and the spectral dimension}

The Gaussian model on $\G$ is defined \cite{hhw} by assigning a
real--valued random variable $\phi_i$ to each node $i\in G$ and then
prescribing the following probability measure 
\begin{equation}
	d\mu_r[\phi] = {1\over Z_r} 
	\exp\left[- \sum_{i,j \in S_{o,r}}\phi_i (\bds L + 
	m^2\,\bds\eta)_{ij}\phi_j \right] \prod_{i\in S_{o,r}} d\phi_i
\end{equation}
for the collection $\phi=\{\phi_i\,;\,i\in S_{o,r}\}$. Here $Z_r$ is
the proper normalization factor, $m>0$ is a free parameter and
$\bds\eta$ is the diagonal matrix with elements
$\eta_{ij}=\eta_i\delta_{ij}$ with the real numbers $\eta_i$ positive
definite and uniformly bounded throughout $G$ (that is
$0<\eta_{\mathrm{min}}\le \eta_i \le \eta_{\mathrm{max}}$, $\forall i\in\G$).

The thermodynamic limit is achieved by letting $r\to\infty$ and
defines a Gaussian measure over the entire $\phi=\{\phi_i\,;\,i\in
G\}$ which does not depend on the centre of the Van Hove sphere $o$ 
\cite{rimtim}. The covariance of this Gaussian process reads
\begin{equation}
\langle \phi_i \phi_j \rangle \equiv  C_{ij}(m^2) = 
(\bds L + m^2 \bds\eta)^{-1}_{ij}
\label{gauss2}
\end{equation}
and hence it satisfies by definition the Schwinger--Dyson (SD) equation 
\begin{equation}
\label{eq:SD}
	(J_i + m^2\eta_i) C_{ij}(m^2) - \sum_{k\in G} J_{ik} C_{kj}(m^2) 
	= \delta_{ij}
\end{equation} 
Setting
\begin{equation}
	C_{ij} = {(1-W)^{-1}_{ij} \over {J_i + m^2 \eta_i}} ,~~~
	W_{ij} = {J_{ij}\over {J_j + m^2 \eta_j}} 
\end{equation}
one obtains the standard connection with the random walk (RW) over
$\G$ \cite{hhw}:
\begin{equation}
\label{eq:RW}
	(1-W)^{-1}_{ij} \,=\,\sum_{t=0}^\infty \, (W^t)_{ij} = 
	\sum_{\gamma:\,i\leftarrow j} W[\gamma]
\end{equation}
where the last sum runs over all paths from $j$ to $i$, each weighted
by the product along the path of the one--step probabilities in $W$:
\begin{equation}
	\gamma = (i,k_{t-1},\ldots,k_2,k_1,j) \Longrightarrow
	W[\gamma] = W_{ik_{t-1}} W_{k_{t-1}k_{t-2}},\ldots,W_{k_2k_1}W_{k_1j}
\end{equation}
Notice that, as long as $m>0$, we have $\sum_i (W^t)_{ij}<1$ for any
$t$, namely the walker has a nonzero death probability. This implies
that $C_{ij}$ is a smooth functions of $m^2$ for $m\ge\epsilon>0$. In
the limit $m\to 0$ the walker never dies and the sum over paths in
eq. (\ref{eq:RW}) is dominated by the infinitely long paths which sample
the large scale structure of the entire graph (``large scale'' refers
here to the metric induced by the chemical distance alone). This
typically reflects itself into a singularity of $C_{ij}$ at $m=0$ whose
nature does not depend on the detailed form of $J_{ij}$ or $\eta_i$,
as long these stay uniformly positive and bounded.

Of particular importance is the leading singular infrared behaviour, as $m^2\to
0$, of the average $[C(m^2)]_G$ of $C_{ii}(m^2)$, which is a positive
definite quantity,  over all points
$i$ of the graph ${\cal G}$, which we may write in general as
\begin{equation}
	{\rm Sing}\, [C(m^2)]_G  \sim c (m^2)^{{\bar d}/2-1} 
\label{leadsing}
\end{equation}
The parameter ${\bar d}$ is called the spectral dimension of the graph
$\G$ and on regular lattices it coincides with the usual Euclidean
dimension. Henceforth we shall call {\em spectral weight} the
coefficient $c$ in eq. (\ref{leadsing}). The name {\it spectral
dimension} is related to the behaviour of the spectral density $\rho(l)$
of low-lying eigenvalues of the Laplacian $\bds L$; indeed it can be
shown \cite{debole} that $\rho(l)$ scales as a power of $\l$ for $\l\to0$,
that is $\rho(l)\sim l^{~{\bar d}/2 -1}$.

\section{Large scale inhomogeneity: homogeneity classes and spectral classes}

In the study of statistical models one often has to deal with the
average $[C(m^2)]_A$ of $C_{ii}(m^2)$ over a generic positive measure
subset $A \subset G$ and in particular one has to consider the leading
singular behaviour of $[C(m^2)]_A$ as $m^2 \to 0$.  On regular lattices
this singular behaviour is independent of $A$ and it actually coincides
with that obtained averaging over all points of ${\cal G}$:
\begin{equation}
{\rm Sing}\,[C(m^2)]_A = {\rm Sing}\, [C(m^2)]_G 
\;,\quad \forall ~A\subset G \;,\quad |A|>0
\label{hom}
\end{equation}
This property arises from the large scale homogeneity of regular
lattices due to translation invariance. On graphs, where translation
invariance is lost, this property can still hold if the inhomogeneity
is limited to finite scales. More generally it may happen that
inhomogeneity extends to large scales and the singular parts of
eq. (\ref{hom}) are different on different subsets.  However we will
prove that such subsets must satisfy very strong topological
constraints: a large--scale inhomogeneous graph always consists of
homogeneous parts joined together by a zero--measure set of links.
Therefore the splitting of infrared behaviour always corresponds to a
macroscopically evident inhomogeneity of the graph.

In this section we will give a rigorous formulation of these
statements through the following steps.
\begin{itemize}
\item 
Let us suppose that the graph $\G$ has indeed a large--scale
inhomogeneity that manifests itself through the existence of at least
one nonzero--measure subset $A \subset G$ such that, as $m^2 \to 0$,
\begin{equation}
{\rm Sing}\, [C(m^2)]_A  \sim c_{A} (m^2)^{{\bar d}_A/2-1}
\label{singa} 
\end{equation}
with ${\bar d}_A \neq \bar d$. 
\item
We then define $M\subset G$ to be a {\em maximally homogeneous} (or
more briefly {\em maximal}) subset with respect to ${\bar d}_A$
whenever
\begin{enumerate}
\item 
$|M\,\cap A| > 0$
\item
${\rm Sing}\, [C(m^2)]_M \sim c_{M} (m^2)^{{\bar d}_M/2-1}$, with ${\bar d}_M
= \bar d_A$.
\item
For any nonzero--measure subset $B\subset M$ we have ${\bar d}_B={\bar d}_M$.
\item
There exists no $B \supset M$ such that ${\bar d}_B={\bar d}_M$ and $|B|> |M|$.
\end{enumerate}
By this definition it follows that the set of all maximal subsets with 
respect to ${\bar d}_M$ coincides with the equivalence class
$\{M\}$ and we will call it the {\it homogeneity class} of ${\bar d}_M$. 

\item 
Next we prove the\\ 
{\bf Theorem 1:} The subgraphs $\M$ and its
complement ${\bar \M}$ are infinitely separated, i.e. their minimal
distance $\underline{D}(\M,{\bar \M})$ is infinite and they define a
separable partition of $\G$. Since this separability is induced by
the spectral properties embodied by the spectral dimension, we call
this a spectral partition (of order two) of $\G$. 

\item
Finally we consider a Gaussian model on the graph $\M$ 
showing that, from the infinite separability 
of $\M$ and ${\bar \M}$ the spectral dimension of $\M$ is ${\bar d}_M$.
Therefore, $\bar d_M$ is a property of the graph 
$\M$ and defines a {\it spectral class}. This chain of arguments may now be
applied to ${\bar \M}$, splitting off a new spectral class if ${\bar \M}$
has a large scale inhomogeneity of the type given above. The process
can be repeated until necessary, yielding a complete spectral partition
of the original graph $\G$ into spectral classes.
\end{itemize}
\begin{flushleft} 
{\bf Proof of Theorem 1:} 

Let us suppose ad absurdum that $\underline{D}(\M,{\bar \M})=1$ and
therefore that there exists a nonzero--measure subset ${\bar M}' \subset
{\bar M}$ such that $\underline{D}(\M,{\bar \M}')=1$. From the
maximality of $M$ it follows that ${\bar d}_M \neq {\bar d}_{{\bar
M}'}$. Let us consider the random walk representation (\ref{eq:RW}) of
$C_{ii}(m^2)$ with $i \in {\bar M}'$:
\begin{equation}
	C_{ii}(m^2) = {1\over {J_i + m^2 \eta_i}} 
	\sum_{\gamma:\,i\leftarrow i} W[\gamma]
\label{sumpath}
\end{equation}
Next consider a site $k \in M$ whose distance from $i$ is $1$. This
site exists from the hypothesis $\underline{D}(\M,{\bar \M}')=1$. Then,
from the sum over paths in the left hand side of (\ref{sumpath}) let
us retain only the paths containing $k$. Then, from the boundedness
and positivity of $J_{ij}$ and $\eta_i$ one gets:
\begin{equation}
	C_{ii}(m^2) \ge  {C_{kk}(m^2)\over {J_{max} + m^2 \eta_{max}}}
\label{sumpath1}
\end{equation} 
Averaging over $M$ and then over ${\bar M}'$ we get:
\begin{equation}
	[C(m^2)]_{{\bar M}'} \ge  K \, [C(m^2)]_M
\end{equation} 
where $K$ is a positive constant. Now, taking $m^2 \to 0$ and using
the asymptotic expression for $[C(m^2)]$ given in (\ref{leadsing}) we
obtain
\begin{equation}
(m^2)^{{\bar d}_{{\bar M}'}/2-1} \ge K' \,(m^2)^{{\bar d}_{M}/2-1} 
\end{equation}
Since this argument applies equally well with $\M$ and ${\bar \M}$
interchanged, one gets:
\begin{equation}
(m^2)^{{\bar d}_M/2-1} \ge K'' \,(m^2)^{{\bar d}_{{\bar M}'}/2-1} 
\end{equation}
which gives ${\bar d}_M = {\bar d}_{{\bar M}'}$ contradicting 
the hypothesis. Therefore $\underline{D}(\M,{\bar \M})=\infty$ and $\M$
and ${\bar \M}$ must be infinitely separated.
\end{flushleft} 

The infinite separability of $\M$ and $\bar \M$ implies that the two
subgraphs can be disconnected by cutting a zero--measure set of links.
This very peculiar property implies {\em thermodynamic independence},
that is the decoupling, in the thermodynamic limit, of a model
defined on the whole graph $\G$ into two models defined independently
on on $\M$ and $\bar \M$ \cite{rimtim}. 

This applies in particular to the Gaussian model, so that the two
averages of $C_{ii}(m^2)$ on $\M$ and $\bar \M$ are independent
quantities, each satisfying a relation like eq. (\ref{leadsing}) with
two distinct spectral dimensions.  Most importantly, to any
nonzero--measure subset of $\M$ there corresponds by construction the
same spectral dimension $\bar d$ of $\M$. We can say then that $\bar
d$ is a universal property of $\M$.

\section{Spectral weights and Subclasses of spectral classes}

In the singular behaviour of $[C(m^2)]$, inhomogeneity at large scale
can appear also in the coefficient of the leading infrared part
(\ref{leadsing}). However, following the same steps as the previous
section, we will show that once again a splitting in the value of the
coefficient corresponds to a macroscopic inhomogeneity of the graph
and that a macroscopically homogeneous graph is indeed characterized
by universal ${\bar d}$ {\em and} $c$. Actually in this case the
proof is subtler and requires some further mathematical steps. 

We first define the spectral {\em subclasses} of a given spectral
class by looking at the spectral weight $c_A$, proceeding along steps
similar to those followed above.
\begin{itemize}
\item 
Let us suppose that, for a given graph $\G$ belonging to the spectral
class characterized by $\bar d$, there exists at least one
nonzero--measure subset $A \subset G$ such that, as $m^2 \to 0$,
\begin{equation}
{\rm Sing}\, [C(m^2)]_A  \sim c_A (m^2)^{{\bar d}/2-1}
\label{singcoeff} 
\end{equation}
with $c_A \neq  c$, with $c$ given as in eq. (\ref{leadsing}). 
\item
Then we say that a nonzero--measure subset $M \subset G$, which
certainly is maximal w.r.t. $\bar d$, due to its universality, is
maximal also w.r.t. $c_A$ whenever
\begin{enumerate}
\item 
$|M\,\cap A| > 0$
\item
${\rm Sing}\, [C(m^2)]_M \sim c_M (m^2)^{{\bar d}/2-1}$, with $c_M
= c_A$.
\item
For any nonzero--measure subset $B\subset M$ we have $c_B=c_M$.
\item
There exists no $B \supset M$ such that $c_B=c_M$ and $|B|> |M|$.
\end{enumerate}
By this definition it follows that the set of all maximal subsets with 
respect to $c_M$ coincides with the equivalence class
$\{M\}$ and we will call it the {\it homogeneity subclass} of spectral
weight $c_M$. 

\item
We then prove the\\
{\bf Theorem 2:} The subgraphs $\M$ and its complement
${\bar \M}$  are infinitely separated and define a spectral
partition of $\G$.
\item
Following the same steps as the previous section, we then 
consider a Gaussian model on the graph ${\cal M}$ showing that, 
from the infinite separability of $\M$ and ${\bar \M}$, 
the coefficient of ${\rm Sing}\,[C(m^2)]_M$ 
is $c_M$. Therefore we can say that $c_M$ is a universal
property of the graph $\M$ and defines a {\it spectral subclass}
separated from the rest.
\end{itemize}

\begin{flushleft} 
{\bf Proof of Theorem 2:} 

To prove this theorem we first need the following lemma: \\

{\bf Lemma}: Within a given spectral subclass, for any subset $A$ of
the subclass, the asymptotic form of $[C(m^2)]_A$ is invariant under
pre--averaging over any normalized point distribution with
nonzero--measure support. In other words, if we define
\begin{equation}
	[C(m^2)]_{A,\alpha} = {[\alpha\,C(m^2)]_A\over [\alpha]_A }
\end{equation}
where $\alpha_i > 0$ on a subset of $A$ with nonzero measure, then again 
\begin{equation}
	{\rm Sing}\,[C(m^2)]_{A,\alpha} \sim c_A (m^2)^{{\bar d}/2-1} 
\end{equation}
with no dependence at all for $c_A$ and $\bar d$ on the distribution 
$\alpha=\{\alpha_i\,;\;i\in A\}$. The proof of this statement is
elementary: we define the quantities 
\begin{equation}
	f_i = (m^2)^{-\bar d/2+1} C_{ii}(m^2) - c_A
\end{equation}
Then, by construction, for any $\epsilon>0$ there exist a $\delta>0$ such
that we have $\left| [f]_A \right| <\epsilon$
as soon as $m^2<\delta$. Hence we also have
\begin{equation}
	\bigl| [\alpha f]_A \bigr| < 
	\bigl(\sup_{i\in A} \alpha_i\bigr) \bigl| [f]_A \bigr|
	< \bigl(\sup_{i\in A} \alpha_i \bigr) \epsilon
\end{equation}
which immediately implies our assertion. 
\end{flushleft}

Now we can prove Theorem 2: 
\begin{flushleft}

Let us suppose ad absurdum that $\underline{D}(\M,{\bar \M})=1$ and
therefore that it exists a nonzero--measure subset ${\bar M}' \subset
{\bar M}$ such that $\underline{D}(\M,{\bar \M}')=1$. From the
maximality of $M$ it follows that $c_M \neq c_{{\bar
M}'}$. 

The following proof is given only for $\bar d<4$, owing to brevity
and physical requirements. Indeed a real structure has necessarily a
dimension $\bar d \le3$; moreover, from a purely theoretical point of
view, the class of models we have in mind, with site variables and
link interactions, typically have $4$ as an upper critical dimension for
the scaling behaviour.   

Let us consider first the case of a spectral class where $[C(m^2)]_G$
diverges when $m^2 \to 0$, that is such that $\bar d<2$.
The Schwinger-Dyson equation for $C_{ii}[m^2]$ reads:
\begin{equation}
	(J_i + m^2\eta_i) C_{ii}(m^2) - \sum_{k\in\G} J_{ik} C_{ki}(m^2) 
	= 1
\label{schdy}
\end{equation} 
Averaging equation (\ref{schdy}) over $M$, we obtain the relation
\begin{equation}
\label{eq:JCetc}
	[J\,C]_M  + m^2 ~ [\eta\,C]_M - [J \cdot C]_M = |M|
\end{equation}
where $(J\,C)_i \equiv J_i C_{ii}$, $(\eta\,C)_i \equiv \eta_i C_{ii}$
and $(J \cdot C)_i = \sum_k J_{ik}C_{ki}$. We then divide by
$[J\,C]_M$ and let $m^2 \to 0$. Due to the divergence
of $[J\,C]_M$ we have that, for any $\epsilon>0$ there exists a
$\delta>0$ such that, as soon as $m<\delta$,
\begin{equation}
	1-\epsilon \le {{[J \cdot C]_M}\over{[J\,C]_M}}
\end{equation}                              
Next we set 
\begin{equation}
	J_{{\bar M}',i} = \sum_{k\in {\bar M}'} J_{ik} \;, \quad 
	(J\cdot C)_{{\bar M}',i} = \sum_{k \in {\bar M}'} J_{ik}C_{ki}  
\end{equation}
and use the positivity of $C_{ii}-C_{ik}$ \cite{hhw} to push the above 
inequality to
\begin{equation}
	1 - \epsilon \le 1 - {{[J_{{\bar M}'}\,C]_M}\over{[J\,C]_M}} 
	   + {{[(J\cdot C)_{{\bar M}'}]_M}\over{[J\,C]_M}}  
\end{equation}
which yields
\begin{equation}
      \lim_{m^2 \to 0} {{[(J\cdot C)_{{\bar M}'}]_M}
      \over{[J_{{\bar M}'}\,C]_M}} =1 
\end{equation}
Owing to the symmetry of $\underline{D}(\M,{\bar \M}')$, we may
repeat the above steps with $M$ and ${\bar M}'$ interchanged. Since
the symmetry of $J_{ij}$ and $C_{ij}$ implies $[(J\cdot C)_{{\bar
M}'}]_M=[(J\cdot C)_M]_{{\bar M}'}$, we finally obtain
\begin{equation}\label{MM'1}
	\lim_{m^2 \to 0} { {[J_{{\bar M}'}\, C]_M}\over{[J_M\,C]_{{\bar M}'}}} 
        = 1 
\end{equation}
At this stage we apply the lemma given above with $\alpha$ identified
with $J_{{\bar M}'}$ or $J_M$, namely
\begin{equation}
	[J_{{\bar M}'}\,C]_M \sim c_M\,
        [J_{{\bar M}'}]_M\,(m^2)^{{\bar d}/2-1}\;,\quad
	[J_M\,C]_{{\bar M}'} \sim c_{{\bar M}'}\,
        [J_M]_{{\bar M}'}\,(m^2)^{{\bar d}/2-1}		
\end{equation}
But $[J_{{\bar M}'}]_M = [J_M]_{{\bar M}'}$ so that eq. (\ref{MM'1})
implies $c_M=c_{{\bar M}'}$, contradicting our initial hypothesis that
$\underline{D}(\M,{\bar \M})=1$ with $M$ maximal. Hence necessarily
$\underline{D}(\M,{\bar\M})=\infty$, proving our assertion.
\vskip 0.3truecm
Let us now consider a spectral class where $C(m^2)_G$ 
does not diverge in the limit $m^2 \to 0$ while its first derivative
with respect to $m^2$, $C'(m^2)_G$, diverges in the same limit.
This is the case of a spectral class characterized by a spectral dimension
$2 < {\bar d} <4$, where:
\begin{equation}
	[C'(m^2)]_{M,\alpha} = {[\alpha\,C'(m^2)]_M\over [\alpha]_M} \sim
        - ({\bar d}/2-1)~ c_M ~(m^2)^{{\bar d}/2-2} \;,\quad m^2 \to 0
\label{divder}
\end{equation}
Taking the first derivative with respect to $m^2$ in the Schwinger-Dyson
equation (\ref{schdy}), we obtain: 
\begin{equation}
	\eta_i C_{ii}(m^2) + m^2 \eta_i C'_{ii}(m^2) =
         \sum_{k\in\G} J_{ik} [C'_{ki}(m^2)-C'_{ii}(m^2)] 
\label{schdyder}
\end{equation} 
which can be averaged over $M$ giving
\begin{equation}
	[\eta~C]_M + m^2 [\eta~C']_M = [J \cdot C']_M - [J~C']_M 
\label{schdydermed}
\end{equation}

Together with eq. (\ref{divder}), this implies
\begin{equation}
      \lim_{m^2 \to 0} (m^2)^{2-{\bar d}/2} ([J \cdot C']_M - [J~C']_M ) = 0^+
\end{equation}
that is,  for any $\epsilon>0$ there exists a
$\delta>0$ such that, as soon as $m^2 <\delta$
\begin{equation}
	0 < \xi ([J \cdot C']_M - [J~C']_M ) < \epsilon
\end{equation}  
with $\xi ~= ~(m^2)^{2-{\bar d}/2}$. This can be rewritten as
\begin{equation}
	0 < [(J \cdot C')_M]_M - [J_M~C']_M  + 
[(J \cdot C')_{\bar M}]_M - [J_{\bar M}~C']_M < \xi^{-1}~\epsilon
\end{equation}  
Now, since $C'_{ij} \equiv - \sum_k \eta_k C_{ik} C_{kj}$ are the
elements of a negative semi--definite matrix, one has
that $[(J \cdot C')_M]_M - [J_M~C']_M >0$. Therefore
\begin{equation}
	0 \le [(J \cdot C')_{\bar M}]_M - [J_{{\bar M}'}~C']_M 
          < \xi^{-1}~\epsilon
\end{equation}  
Again owing to the symmetry of $\underline{D}(\M,{\bar \M}')$, the
previous steps can be repeated with $M$ and ${\bar M}$ interchanged,
leading to:
\begin{equation}
	0 \le [(J \cdot C')_M]_{\bar M} - [J_M~C']_{{\bar M}'} 
          < \xi^{-1}~\epsilon
\end{equation}  
Since $[(J \cdot C')_{\bar M}]_M = [(J \cdot C')_M]_{\bar M}$, 
these two relations imply:
\begin{equation}
        0 \le | [J_{{\bar M}'}~C']_M - [J_M~C']_{{\bar M}'}| 
         <  \xi^{-1}~\epsilon
\label{dismod}
\end{equation}
Eq. (\ref{divder}) entails in the limit $m^2\to 0$:
\begin{equation}
	[J_{{\bar M}'}~C']_M \sim
        - ({\bar d}/2-1)~c_M~[J_{{\bar M}'}]_M~\xi^{-1} \;,\quad 
	[J_M~C']_{{\bar M}'} \sim
        - ({\bar d}/2-1)~c_{{\bar M}'}~[J_M]_{{\bar M}'}~\xi^{-1}
\end{equation}
so that, since $[J_{{\bar M}'}]_M = [J_M]_{{\bar M}'}$ from (\ref{dismod}) 
one obtains $c_M=c_{{\bar M}'}$, which contradicts our hypothesis 
$\underline{D}(\M,{{\bar \M}'})=1$
and therefore proves our assertion $\underline{D}(\M,{\bar \M})=\infty$. 
\end{flushleft} 


\end{document}